\documentclass[aip, linenumbers, reprint, longbibliography]{revtex4-2}

\draft 
\usepackage{graphicx}
\usepackage{subfigure}
\usepackage{ulem}
\usepackage{color}
\usepackage{bm}
\usepackage{comment}
\usepackage{soul}
\usepackage{amsmath,amssymb}

\usepackage{todonotes}
\definecolor{electricpurple}{rgb}{0.75, 0.0, 1.0}

\begin{document}

\nolinenumbers


\preprint{Submitted to}

\title{Spectral flow of a localized mode in elastic media}

\author{M. Miniaci}
\email{marco.miniaci@univ-lille.fr}
\affiliation{CNRS, Univ. Lille, Centrale Lille, Univ. Polytechnique Hauts-de-France, Junia, UMR 8520 -IEMN, F-59000 Lille, France}
\author{F. Allein}%
\affiliation{CNRS, Univ. Lille, Centrale Lille, Univ. Polytechnique Hauts-de-France, Junia, UMR 8520 -IEMN, F-59000 Lille, France}
\author{R. K. Pal}%
\affiliation{Mechanical and Nuclear Engineering Department, Kansas State University, Manhattan, Kansas, USA}

\date{\today}


\maketitle 



\noindent\textbf{The introduction of structural defects in otherwise periodic media is well known to grant exceptional space control and localization of waves in various physical fields, including elasticity.
\\
Despite the variety of designs proposed so far, most of the approaches derive from contextual modifications that do not translate into a design paradigm due to the lack of a general theory. 
Few exceptions include designs endowed with topological dispersion bands, which, however, require changes over substantial portions of the structure.
\\
To overcome these limitations, here we introduce a new rationale based on real-space topology to achieve localized modes in continuous elastic media.
We theoretically predict and experimentally demonstrate the spectral flow of a localized mode across a bulk frequency gap by modulating a single structural parameter at any chosen location in the structure.
\\
The simplicity and generality of this approach opens new avenues in designing wave-based devices for energy localization and control.}

\section*{\noindent \textbf{Introduction}}
\noindent The quest for media capable of exceptional wave localization has attracted increasing interest in many research fields, owing to its promise of fostering new functionalities, such as defect-immune and scattering-free wave propagation~\cite{hasan2010colloquium}, unprecedented potential for energy harvesting applications~\cite{aouali2020efficient}, object cloaking~\cite{stenger2012experiments} and enhanced energy transport~\cite{stutzer2018photonic}, to cite a few.
For this reason, wave localization has been extensively studied for over a century in various physical domains, including electromagnetism, elasticity and acoustics.
In particular, the fundamental interplay between local properties (material composition and geometrical architecture) and structural defects has been proven to lead to the emergence of localized modes~\cite{flach2008discrete} offering intriguing possibilities for spatial wave control.
\\
Localized modes can be broadly classified through the symmetry of the structure hosting the wave propagation, i.e., (i) random / disordered~\cite{lagendijk2009fifty,sheng1990scattering, weaver1990anderson, ye1992observation, hu2008localization}, (ii) quasi-periodic~\cite{zhang2012defect, hladky2013acoustic, chen2015twisting, povzar2018isolated} and (iii) periodic media with the introduction of defects~\cite{khelif2003trapping, benchabane2005interaction, zheng2019granular}.
In this context, continuous elastic media offer a rich playground in such a quest because of their fourth order tensor-based physics coupling longitudinal, shear and flexural deformations~\cite{graff2012wave}.
It has been shown that elastic localized modes can be achieved by breaking specific symmetries by adding or removing inclusions in the unit cell, by varying their size and shape in order to create point~\cite{khelif2003trapping, marchal2012dynamics, yang2015high, jo2020graded, jo2020designing} or line defects~\cite{sigalas1998defect, benchabane2015guidance, wang2018guiding}.
\\
However, despite the variety of designs proposed so far, all these approaches are based on ad-hoc modifications that do not translate into a general design paradigm.
Hence they do not allow a systematic prediction, a priori, of the presence / absence of a localized mode under a structural modification.
Another limitation of these approaches is that the induced localized modes are extremely sensitive to the presence of additional defects in the structure, meaning a high risk of uncontrolled frequency shifts of the modes.  
\\
The recent introduction of topological protection in elasticity has opened new possibilities for a more systematic design procedure for achieving localized modes insensitive to defects~\cite{mousavi2015topologically, susstrunk2015observation, nash2015topological, miniaci2018experimental, zhang2018topological} because endowed with nontrivial, or topologically protected, dispersion bands.
However, solutions explored so far require material / geometrical modifications over the entire or substantial portions of the structure~\cite{ma2019topological}, due to the fact that such modes arise at the interface between two domains with distinct topological properties characterized by different topological invariants, such as the Zak phase or Chern number~\cite{miniaci2021design}.
In this context, the recent introduction of the concept of fragile topological phases~\cite{po2018fragile, peri2020experimental} has challenged this notion by demonstrating the presence of topological modes even in the  absence of  nontrivial invariants.
A spectral flow of a family of localized modes arising across an interface under twisted boundary conditions~\cite{song2020twisted}, i.e., as specific physical parameters are smoothly changed, has been demonstrated.
However, these observations are limited to pressure acoustic waves~\cite{chen2021nonlocal, peri2020experimental} (where the propagation is described through a scalar field potential), electronic charges localizing at corners and dislocation cores due to filing anomaly-induced topological effects~\cite{benalcazar2019quantization}, and photonic lattices~\cite{peterson2021trapped}.
On the contrary, the spectral flow of localized modes in continuous elastic systems has remained elusive so far, due to their unique tensor-based nature of its wave equations~\cite{graff2012wave} implying high modal density and their tendency to hybridize under structural modifications.
\\
In this Letter, using polymeric 3D-printed metamaterials, we report for the first time the experimental observation of a family of localized modes spanning a frequency bandgap between two bulk dispersion bands. 
We observe the spectral flow of this mode using 10 different samples that accommodate a smooth local modulation of the stiffness of a single unit cell.
The observed behaviour is accompanied by a thoroughly interpretation based on a theoretical model allowing to fully predict, a priori, the presence or absence of the localized mode as a function of the modulated mechanical parameters.
The key idea is to determine the number of natural frequencies in each dispersion band for finite structures with defects by solely exploiting real-space topological properties.
In contrast to prior works on localized interface modes, the structure on both sides of the modulated unit cell remains here identical.
\\
The simplicity and generality of the proposed approach, clearly showing that the centre of the localized mode can be arbitrary chosen within the original structure, thus suppressing the need of substantial modifications of the initial structure, as required by traditional topological protection approaches, may open new avenues in designing wave-based devices for elastic energy localization and control.

\section*{\noindent \textbf{Results}}
\noindent \textbf{Modal spectral flow through stiffness modulation}.
The confirmation of the above idea is obtained by measuring the frequency response function of the axial displacements of periodic one-dimensional ($1D$) mass-spring systems, a schematic representation of which is reported in Fig.~\ref{Fig1}A. 
The rectangle shaded in light yellow represents the unit cell of the chain consisting of two identical masses (green dots) and two springs of alternating stiffness $k$ and $\delta$, respectively.
A defect spring (highlighted in red) located in the middle of the chain is characterized by a stiffness modulation parameter $\lambda$.
Figure~\ref{Fig1}B reports a $3D$ rendering of the printed structures manufactured to realize the elastic analogue of the above described discrete system.
The unit cell (reported in yellow for the sake of clarity) can be divided into three regions: a thick one with square cross-section, corresponding to the masses of the discrete chain (and referred as \textit{masses} in what follows), and two thin beams with circular cross-section of different radii, corresponding to the springs of stiffness $k$ and $\delta$ (and referred as \textit{springs} in what follows).
The axial stiffness of each beam is $EA/L$ to first order, being $E$, $A$ and $L$ the material Young's modulus, the cross-section area and length of the beam, respectively (see Methods for further details).
\\
The first step towards the observation of a topologically protected modal spectral flow is to realize a complete bandgap for all the possible elastic polarizations, namely bending, shear, torsional and axial modes (see Supplementary Note 1~\cite{SI} for further details).
This is done by properly choosing the geometrical parameters of the unit cell under investigation, when $k \ne \delta$.
In our case, a gap between 5840 Hz and 8280 Hz is opened when $k = 2 \delta$ (see Methods).
In the second step, a proper local modulation of the spring stiffness of rigidity $k$ of the unit cell in a finite system (at an arbitrary location) is introduced to induce selective shifts in the natural frequencies of the eigenmode closest to the bandgap.
The stiffness modulation is achieved through a gradual variation of the radius $\varphi$ of the central beam of the finite chain (highlighted by grading colors shading from white to dark blue in Fig.~\ref{Fig1}B).
Ten stiffness modulations are introduced (as $\lambda \cdot k$, with $\lambda \in [0.1,1]$) and enumerated as $\#1 - \#10$, corresponding to $\lambda = 1$ and $\lambda = 0.1$, respectively (supplementary Note 2~\cite{SI} reports a photograph of the manufactured samples).
\\
To unequivocally distinguish between a trivial stiffness alteration and a topologically protected one leading to a modal spectral flow, two distinct cases are considered, namely (i) $k > \delta$ and (ii) $k < \delta$.
The transmissibility of the two classes of systems is investigated by scanning laser Doppler vibrometry (SLDV) of the samples (longitudinal velocities are excited and acquired - see Methods and Supplementary Note 3~\cite{SI} for further details on the experimental configuration).
The transmissibility is calculated as the ratio of the detected velocity amplitude at scanning points and the imposed one.
Figure~\ref{Fig1}C reports the frequency response function in the $0 - 11$ kHz frequency range for the two study cases ($k > \delta$ in the left panel, and $k < \delta$ in the right panel) when 10 different values of $\lambda$ are adopted.
In the first case, a spectral flow of the 8$^{\rm{th}}$ mode from the lower to the upper bulk region is clearly observed as energy spots passing across the entire bandgap when $\lambda$ is varied in the $[0.1,1]$ range.
On the contrary, when $k < \delta$, no crossing is observed as the parameter $\lambda$ is changed.
This provides a direct observation of the aforemetioned modal spectral flow.
The experimental results are perfectly supported by analytical models reported as overlaid white squares (natural frequencies of the axial modes of the specimens) and by time domain numerical calculations (see Supplementary Note 4~\cite{SI}).
It is worth mentioning here that rigid longitudinal translation modes at zero frequency are filtered out in the experimental measurements through a high-pass band filter.
\\

\noindent \textbf{Eigenvector reconstruction and localization of the flowing mode}.
When $k > \delta$, a family of modes, whose natural frequency traverses the bandgap as the radius of a single beam is varied, exists. 
To verify that these modes are indeed localized (contrary to the case of $k < \delta$), the 8$^{\rm{th}}$ mode shape is reconstructed (in terms of amplitude and phase) for the two classes of systems. 
\\
Figure~\ref{Fig2}A reports the normalized amplitudes of the displacement of the masses for the 8$^{\rm{th}}$ mode shape of the two chains ($k > \delta$, left panel, and $k < \delta$, right panel) setting the parameter $\lambda$ to 1 (highlighted by green arrows and circles in Fig.~\ref{Fig1}C).
Measured values, reported as red dots, are superimposed on the analytical predictions (blue lines with square markers), and an excellent agreement is found.
In both cases the displacement of the masses spans the entire chain, clearly confirming that the mode can be identified with a Bloch mode lying in the higher and lower bulk bands, as indicated by the green arrows at $\lambda = 1$ in Fig.~\ref{Fig1}B, pointing up or down, respectively.
Each Bloch mode identification can be done via an injective mapping in the corresponding discrete mass-spring chain (see Supplementary Note 5~\cite{SI}).
\\
Figure~\ref{Fig2}B reports the normalized amplitudes of the displacement of the masses for the 8$^{\rm{th}}$ mode shape of the two chains ($k > \delta$, left panel, and $k < \delta$, right panel) now setting the parameter $\lambda$ to 0.5.
In contrast to the previous case, the chain with $k > \delta$ (left panel) clearly presents a mode shape localized at the center of the structure (in correspondence of the unit cell exhibiting the stiffness modulation), with its displacement magnitude rapidly dropping away in the peripheral masses.
On the contrary when $k < \delta$ (right panel) the 8$^{\rm{th}}$ mode shape, still belonging to the (lower) bulk band shows no localization.
Its displacement is maximum at the sample boundaries and is characterized by a slight amplitude decrease at the center of of the structure.
\\
Also in this case, the experimental results are further corroborated by numerical finite element models reporting the deformation of the mode shapes below each subfigure.
Some of the masses are numbered for reference.
The axial displacement of the masses with respect to their equilibrium position is provided (the reference system is set at the center of the chain and odd masses are connected by gray arrows to the abscissa axis of the figure for the sake of clarity) and confirms the existence and the spectral modal flow across the bandgap.
\\

\noindent \textbf{Rationale: spectral counting and topology flow}.
To understand the rationale of the observed spectral flow as $\lambda$ is varied in the $k > \delta$ chain, two discrete chains in the form of rings, as the ones shown in Fig.~\ref{Fig3}A, are considered.
Depending on where the stiffness modulation is applied, i.e., to the spring of stiffness $k$ or $\delta$ (springs highlighted in red), two conditions are possible:
(i) if $\lambda = 1$, both chains reported in Fig.~\ref{Fig3}A reduce to the closed one reported in the left panel of Fig.~\ref{Fig3}B;
(ii) on the contrary, when $\lambda = 0$, the chains reported in the left and right panels of Fig.~\ref{Fig1}A result into the two open chains in the left panels of Figs.~\ref{Fig3}C and D, respectively, depending on whether the $k$ or $\delta$ spring is removed.
Therefore, depending on the stiffness variation of the springs, two classes of chains can be identified: closed (in the form of a ring) or open (one spring is missing) ones.
\\
It is worth noticing here that in both considered cases, open or close finite chains, all the natural frequencies lie on bulk bands (light blue rectangles reported in Figs.~\ref{Fig3}E and~\ref{Fig3}F (refer to Supplementary Note 5~\cite{SI} for additional details).
Counting the number of such modes on each dispersion bulk band lead to different numbers for the different chain configurations.
This difference is the key factor determining the observed spectral flow.
\\
Indeed, when the limit condition of $\lambda = \delta = 0$ is enforced, the considered chains become disjoint units of masses and springs, as shown in each right panel of Figs.~\ref{Fig3}B-D.
Each connected mass-spring subsystem generates eigenmodes with natural frequencies assuming $0$ or $\sqrt{2k/m}$ values.
Hence, counting the number of such disjointed mass pairs gives the corresponding multiplicities of natural frequencies in these limit-case systems.
In particular, both the closed ring of Fig.~\ref{Fig3}B (left panel) and the open ring of Fig.~\ref{Fig3}D (left panel) have $N$ multiples of $\omega = 0$ frequencies when $\lambda = \delta = 0$ (right panels), where $N$ is the number of unconnected mass-spring systems.
On the contrary, the open chain in Fig.~\ref{Fig3}C (left panel) has $N + 1$ multiples of $\omega = 0 $ frequencies and $N - 1$ multiples of $\sqrt{2k/m}$ frequencies, due to the fact that going to the limit of $\lambda = \delta = 0$ two masses are left unconnected.
\\
Now, since all the natural frequencies lie on the dispersion bulk band and the bands do not touch when $k > \delta$, the number of eigenmodes on each band also is expected to remain unaltered for all the family of possible chains with decreasing $\delta$ (up to the limit condition of $\lambda = 0$).
However, if $\lambda$ is varied from $0$ to $1$ for both chains in Fig.~\ref{Fig3}A, a closed chain starting from the two open chains in Figs.~\ref{Fig3}C and~\ref{Fig3}D is obtained, respectively.
In the first case, the number of modes on the lower band must change from $N+1$ to $N$ as we transition from the open to the closed chain.
Since the eigenvalues of the stiffness operator and thus natural frequencies of the chains vary smoothly with this parameter $\lambda$, the only way to achieve this change in number of modes is to have a net spectral flow of one mode from the lower to the upper bulk band (see Fig.~\ref{Fig3}E).
On the other hand, for the case of spring stiffness changing by $\lambda \cdot \delta$, the number of modes remain the same on each band and thus no net spectral flow will happen as $\lambda$ varies from $0$ to $1$ (see Fig.~\ref{Fig3}F).
\\
This reasoning clearly explains the presence (absence) of spectral flow observed in the finite chains in Fig.~\ref{Fig1}C for $k > \delta$ ($k < \delta$).

\section*{\noindent \textbf{Discussion}}
\noindent In conclusion, we have presented a new design strategy to systematically achieve localized modes in continuous elastic media based on real-space topology.
We have demonstrated the possibility of a spectral flow across a bulk frequency gap of a localized mode via thorough experimental measurements.
This is achieved by modulating a single structural parameter at a chosen location of a 3D printed analogue of an elastic mass-spring chain.
The underlying mechanism responsible for such a transition has been explained via a detailed analytical model and corroborated by additional numerical calculations.
\\
This step constitutes a fundamental leap forward in the context of systematic mode localization in elastic media, allowing to control its response without requiring material / geometrical modifications over the entire or substantial portions of the structure.
It presents several advantages over other known techniques, such as (i) only locally varying the unit cell stiffness and (ii) a controlled frequency shift of the spectral flow.
\\
The proposed structure constitutes only a specific example of a new class of materials that can be accessible by applying 3D printing techniques.
We believe these results will open exciting new opportunities in elasticity and acoustics, with the possibility of obtaining novel effects of energy localization and unprecedented degrees of control thanks to the solely local application of stiffness variation, making real-space topology a powerful design strategy of interest in all the fields where vibrations play a crucial role, such as for instance civil, aerospace and mechanical engineering.

\vspace{0.5cm}
\section*{\noindent \textbf{Methods}}
\begin{figure}
\centering
\begin{minipage}[]{1\linewidth}
{\includegraphics[trim=0mm 152mm 72mm 18mm, clip=true, width=1\textwidth]{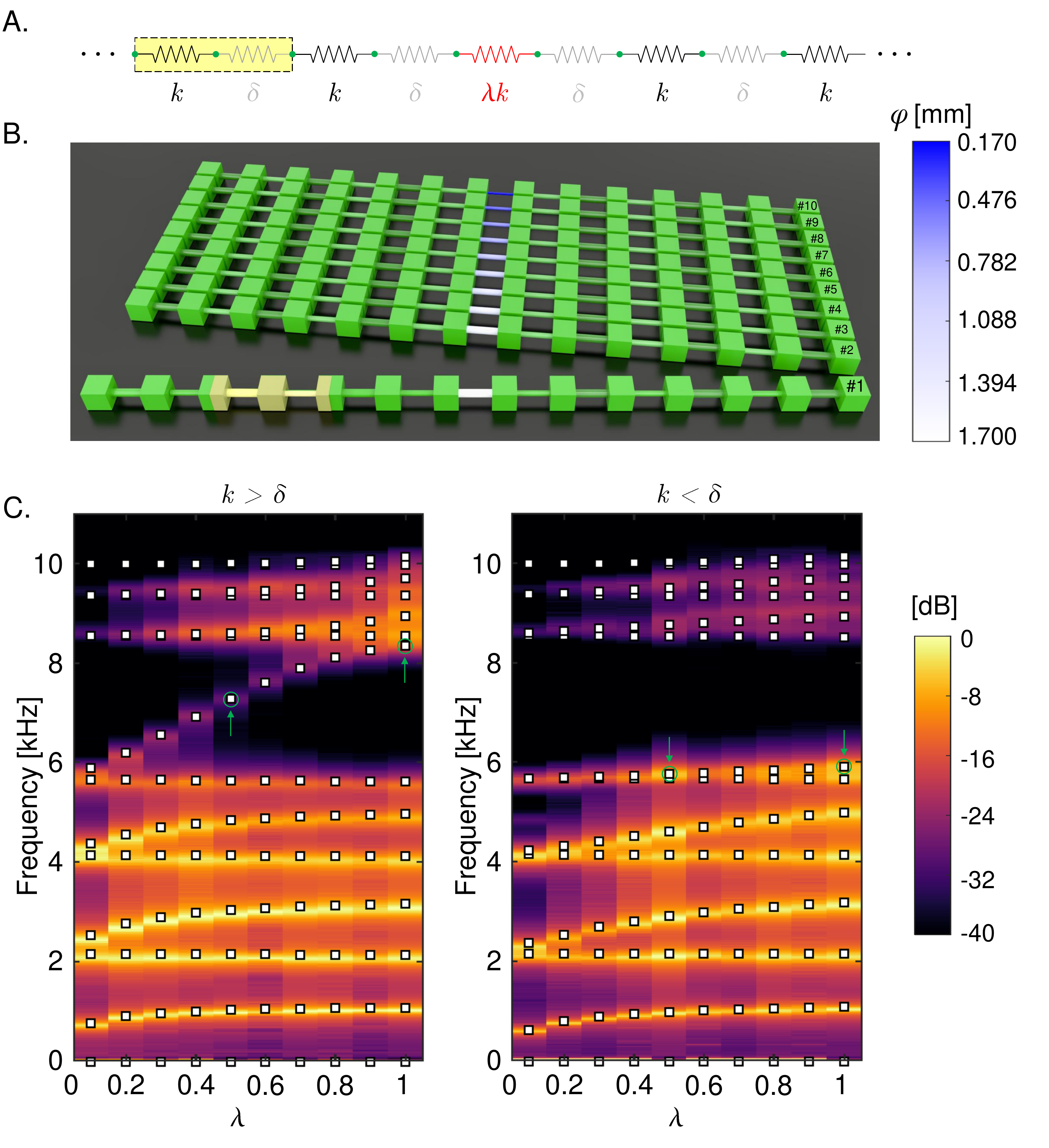}}
\end{minipage}
\caption{\textbf{Observation of topologically protected spectral flow in elasticity}.
\textbf{A}. Schematic representation of a mono-dimensional mass-spring chain.
The unit cell is highlighted as a light yellow rectangle and comprises two masses (green dots) and two springs of stiffness $k$ (in black) and $\delta$ (in grey), respectively.
A defect spring (in red) located in the chain is characterized by a stiffness modulated through the parameter $\lambda$.
\textbf{B}. Three-dimensional rendering of the experimental samples (for the case $k > \delta$).
The stiffness modulation is obtained by gradually varying the radius $\varphi$ of the central beam (highlighted by grading colors going from white to dark blue - see the colormap on the right) connecting its two adjacent masses.
Ten stiffness modulations indicated as $\#1 - \#10$ are considered.
Refer to Methods for further details on the geometrical parameters of the unit cells.
\textbf{C}. Measured frequency response functions (colormap) in the 0 – 11 kHz frequency range for the two classes of elastic chains ($k > \delta$, left panel, and $k < \delta$, right panel) for different values of $\lambda$.
In the first case, a spectral flow of the 8$^{\rm{th}}$ mode from the lower to the upper bulk band is clearly observed (the mode passes across the entire bandgap as $\lambda$ is varied in the $[0.1,1]$ range).
Contrary, in the latter case, no crossing is observed.
The data at each $\lambda$ have been individually normalized with respect to their maximum value.
Overlaid square white dots indicate analytically calculated eigenmodes.
Green arrows indicate the mode shapes that are fully reconstructed and presented in Fig.~\ref{Fig2}.
Elastic waves are excited at the left edge of the specimen and longitudinal displacement is acquired at the opposite edge through a Scanning Laser Doppler Vibrometer.
}
\label{Fig1}
\end{figure}

\begin{figure}
\centering
\begin{minipage}[]{1\linewidth}
{\includegraphics[trim=3mm 160mm 25mm 10mm, clip=true, width=1\textwidth]{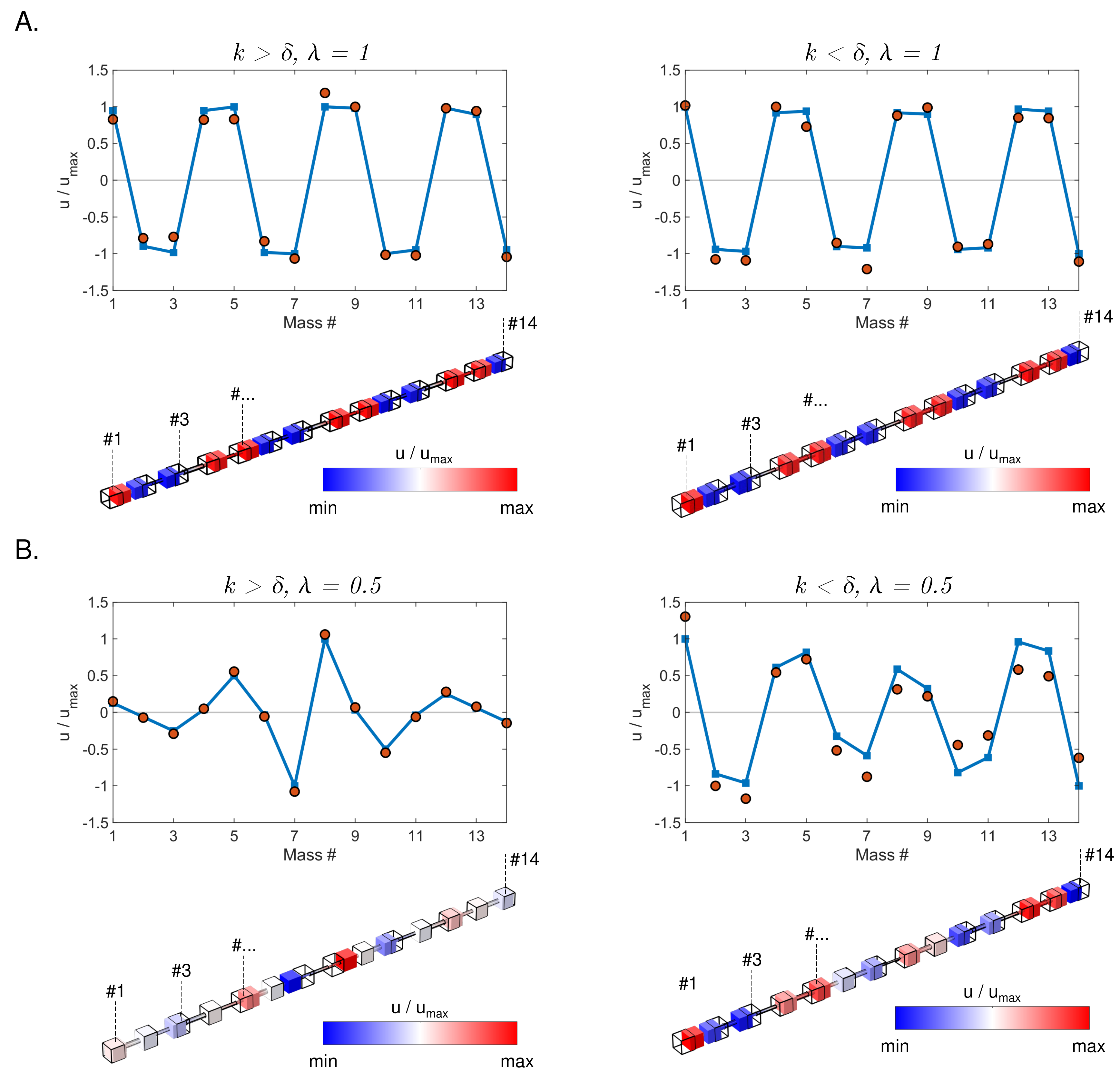}}
\end{minipage}
\caption{\textbf{Eigenvector reconstruction and localization of the flowing mode}.
\textbf{A}. Normalized amplitudes (measured vs. analytical predictions) of the axial displacement of masses for the 8$^{\rm{th}}$ mode shape of the two chains ($k > \delta$, left panel, and $k < \delta$, right panel) setting the parameter $\lambda$ to 1.
In both cases the displacement of the masses spans the entire chain, clearly confirming that the mode can be identified as a Bloch mode lying in a bulk band.
Measurements are reported as red dots and analytical predictions as blue lines with square markers.
\textbf{B}. Normalized amplitudes (measured vs. analytical predictions) of the axial displacement of masses for the 8$^{\rm{th}}$ mode shape of the two chains ($k > \delta$, left panel, and $k < \delta$, right panel) setting the parameter $\lambda$ to 0.5.
The chain with $k > \delta$ (left panel) clearly presents a mode shape localized at the center of the structure, with its displacement magnitude rapidly dropping away in the peripherical masses.
On the contrary when $k < \delta$ (right panel) the 8$^{\rm{th}}$ mode shape, belonging to the (lower) bulk band, shows no localization.
Also in this case, measurements are reported as red dots and analytical predictions as blue lines with square markers.
The insets below each subfigure show the displacement of the masses (deriving from numerical models) with respect to their equilibrium position, with the reference system at the center of the chain.
}
\label{Fig2}
\end{figure}

\begin{figure}
\centering
\begin{minipage}[]{1\linewidth}
{\includegraphics[trim=0mm 0mm 0mm 0mm, clip=true, width=1\textwidth]{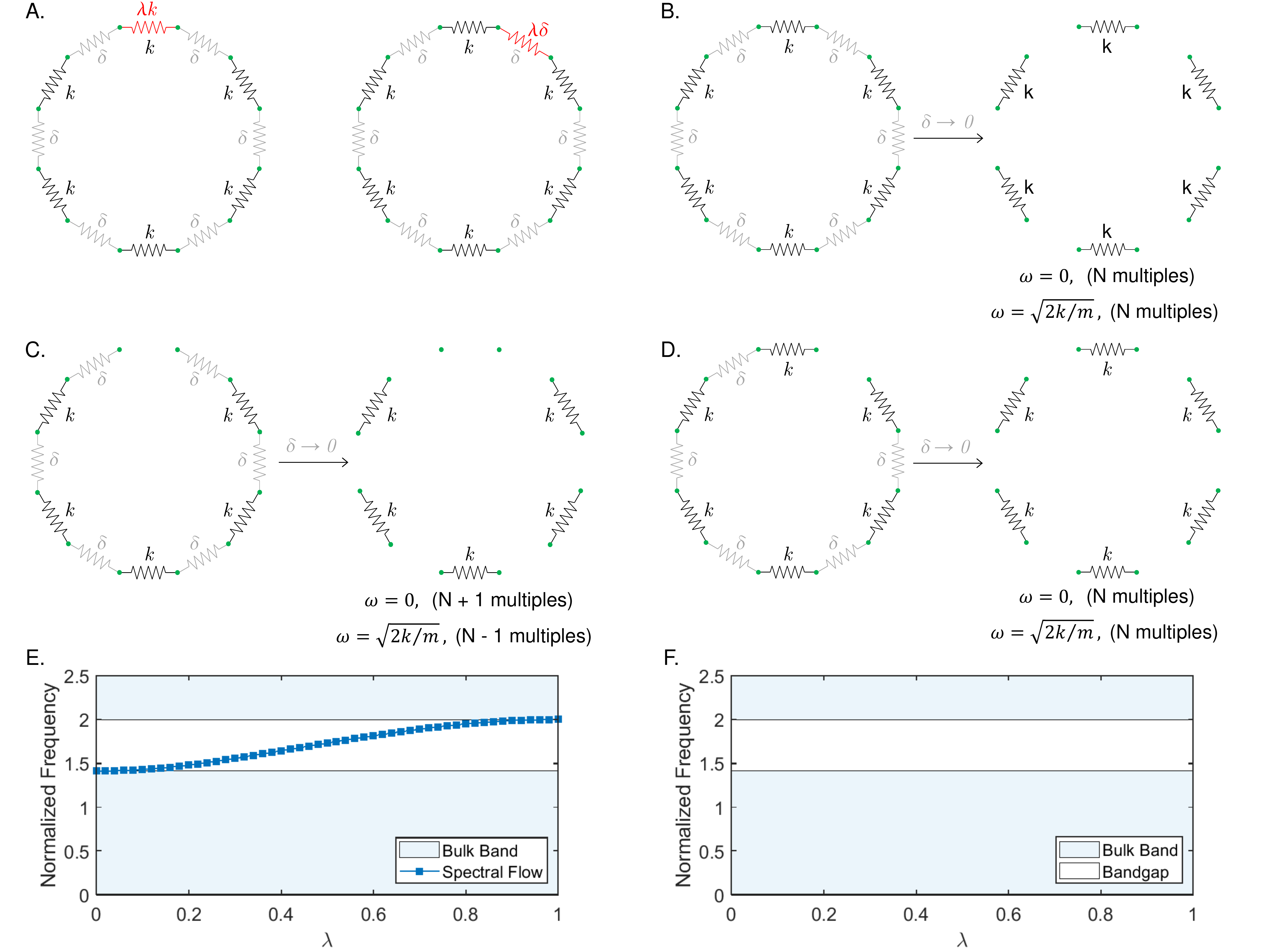}}
\end{minipage}
\caption{\textbf{Theoretical framework explaining the observed spectral flow}.
\textbf{A}. Schematic representation of two $1D$ mass-spring chains arranged in the form of a ring.
The modulation parameter $\lambda$ is here applied to the spring of stiffness $k$ and $\delta$ (springs highlighted in red), respectively.
If $\lambda = 1$, the two chains both reduce to the ring reported in the left panel of subfigure \textbf{B}.
If $\lambda = 0$, two different open chains are obtained by removing a spring with $k$ or $\delta$ stiffness (left panels of subfigures \textbf{C} and \textbf{D}, respectively).
In both cases, the natural frequencies of these systems entirely lie into the bulk bands (frequencies indicated by the light blue rectangles in subfigures \textbf{E} and \textbf{F}).
Taking the lower spring stiffness $\delta$ to 0 (we recall here that $k > \delta$) as a limiting process $(\delta = 0)$, two different sets of disjoint chains are obtained. 
They are characterized by a couple of natural frequencies $\omega$ with multiplicity $N$ (subfigures \textbf{B} and \textbf{D}) or $N + 1$ and $N - 1$ (subfigure \textbf{C}).
As $\lambda$ is varied from 0 to 1, the systems in subfigure \textbf{A} both experience a transition from open to closed chain.
This implies that the number of modes on the lower bulk band changes from $N + 1$ to $N$, transitioning from the open to the close chain.
\textbf{E}. Since the eigenvalues of the stiffness operator, and thus the natural frequencies of the chains, vary smoothly with the parameter $\lambda$, the only way to achieve such a change in number of modes is to have a net spectral flow of one mode from the lower to the upper bulk bands.
\textbf{F}. On the contrary, when the spring stiffness $\delta$ is changed through the parameter $\lambda$, the number of modes remains the same on each band ($N$), thus, implying no net spectral flow.
}
\label{Fig3}
\end{figure}

\small{\noindent\textbf{Simulations}.
Three-dimensional (3D) mode shapes presented in Fig.~\ref{Fig2} are computed via finite element methods using COMSOL Multiphysics software.
The following mechanical parameters are used for the material adopting a linear elastic constitutive law: density $\rho = 1180 \;\text{kg/m}^3$, Young's modulus $E = 2.96$ GPa, and Poisson ratio $\nu = 0.38$.
Domains are meshed by means of three-dimensional 8-node hexahedral quadratic elements of maximum size $L_{FE} = 0.5 $ mm, which is found to provide accurate eigensolutions up to the frequency of interest.
The color-map of the mode shapes reported in Fig.~\ref{Fig2} describes the axial displacement of each mass from its equilibrium configuration.
It varies from negative (blue) to positive (red) and it is normalized with respect to the displacement maximum value.}
\\
\\
\small{\textbf{Sample manufacturing and experimental measurements}.
The specimens, consisting of two classes of chains hosting $14$ masses, are fabricated through additive manufacturing (Stratasys Objet350 Connex3).
Thermoplastic polymer (VERO$^{\rm{TM}}$), with the following nominal properties has been used: density $\rho = 1180 \; \text{kg/m}^3$, Young modulus $E = 2.96$ GPa, and Poisson ratio $\nu = 0.38$.
\\
The geometrical parameters are the following: size of the square mass $a = 10$ mm, length of the beams $L = 15$ mm, radii of the beams corresponding to the springs of stiffness $k$ and $\delta$ in the discrete case, respectively of $r_1 = 1.20 $ mm (k) and $r_2 = 1.20 \cdot \sqrt(2) \simeq 1.70 $ mm ($\delta$) for the case $k < \delta$ and $r_1 = 1.20 \cdot \sqrt(2) \simeq 1.70$ mm (k) and $r_2 = 1.20$ mm ($\delta$) for the case $k > \delta$.
In both cases the radius of the defect is $r_d = r_2 \cdot \sqrt(\lambda)$.
\\
The 10 stiffness modulations for the two classes of chains are obtained through a gradual variation of the radius $\varphi$ of the central beam introduced as $\lambda \cdot k$, with $\lambda \in [0.1,1]$.
\\
The experimental frequency response function presented in Fig.~\ref{Fig1}C is calculated as the ratio of the detected (averaged over 100 times) and the imparted velocity at the acquisition and excitation point, respectively.
Elastic waves are excited impacting manually the left edge of the specimen and longitudinal wave velocity is acquired at the opposite edge through a scanning Laser Doppler Vibrometer (SLDV).
The laser sensitivity was set to $20$ mm$/$s$/$V for all the measurements.
The data at each $\lambda$ has been individually normalized with respect to its maximum value.
\\
A similar procedure is adopted for the experimental mode shape reconstruction reported in Fig.~\ref{Fig2}, and mass $\#7$ was selected as the excitation point (see Supplementary Note 3~\cite{SI} for further details).}


\vspace{0.5cm}
\noindent\textbf{Data availability}.
\noindent The data that support the plots within this paper and other findings of this study are available from the corresponding author upon request.
\\

\noindent\textbf{Acknowledgements}.
\noindent M.M. is funded by the European Union's Horizon 2020 FET Open (Boheme) under grant agreement No. 863179.
R.K.P. is supported by startup funds from Kansas State University and by U.S National Science Foundation Award No. 2027455.
F.A. acknowledges N. Herard, B. Skoropys, M. Coimbra, T. Lata, N. Compton, I. Frankel and N. Boechler for earlier investigation on the experimental realization of a mass-spring system via additive manufacturing.
\\

\noindent\textbf{Author contribution}.
\noindent All authors contributed extensively to the work presented in this paper.
R.K.P conceived the research and developed the theory, M.M and F.A designed and conducted the numerical simulations and experiments.
\\

\noindent\textbf{Additional information}.
\noindent Supplementary information is available in the online version of the paper.
\\

\noindent\textbf{Competing financial interests}.
\noindent The authors declare no competing financial interests.

\onecolumngrid 

\end{document}